\documentclass[12pt]{iopart}
\usepackage{multirow}
\usepackage{graphicx}
\usepackage{dcolumn}
\usepackage{bm}

\begin{document}

\title[]{Lattice coupling and Franck-Condon effects in $K$-edge resonant inelastic X-ray scattering}

\date{}

\author{J. N. Hancock$^{1,2}$, G. Chabot-Couture$^{3}$ and M. Greven$^{3,4}$}

\address{$^{1}$Stanford Synchrotron Radiation Laboratory, Stanford, California 94309, USA}
\address{$^{2}$D\'epartement de Physique de la Mati\`ere Condens\'ee, 24 quai Ernest-Ansermet, Universit\'e de Gen\`eve, Gen\`eve, CH-1211}
\address{$^{3}$Department of Applied Physics, Stanford, California 94305, USA}
\address{$^{4}$School of Physics and Astronomy, University of Minnesota, Minneapolis, Minnesota 55455, USA}

\begin{abstract}
We discuss recent Cu $K$-edge resonant inelastic scattering (RIXS) results for CuB$_2$O$_4$, a model system consisting of electronically-isolated CuO$_4$ plaquettes, which exhibits anomalies in the incident photon energy dependence that are not captured by electron-only models. We show that these anomalous features can be qualitatively described by extending the electron-only considerations to include the lattice degrees of freedom. Our findings have important general implications for the interpretation of $K$-edge RIXS results for electronically higher-dimensional transition metal oxides.
\end{abstract}

\maketitle

\section{Introduction}
Resonant inelastic X-ray scattering can be regarded as a resonant Raman process which results in electronically excited final states reached through intermediate core-hole states of X-ray edges. This species-specific, momentum-tunable probe is enabling the study of Mott-Hubbard physics \cite{abba99, hasan00, lu05, lu06}, Kondo phenomena in rare-earth intermetallics \cite{dallera02}, $d$$\rightarrow$$d$ excitations of transition metal species in complex oxides \cite{ghir04, hanc09} and, more recently, orbital \cite{ulrich09} and magnetic \cite{hill08, braich09} excitations. While this promising technique has led to new insight into the excitation spectra of complex materials, its status as an experimental probe is relatively immature in comparison to the fields of optical or photoemission spectroscopy.

Here we describe the consequences of common considerations \cite{hanc09, vernay08} of the RIXS cross section of a single CuO$_4$ plaquette, resonantly excited through the Cu $K$ (1$s$$\rightarrow$4$p$) X-ray absorption edge. Motivated by recent experimental work for the relatively simple model system CuB$_2$O$_4$, we then point out aspects of experimental data \cite{hanc09, kim04mo, kim04lico} which cannot be explained by electron-only considerations \cite{hanc09, vernay08}. The discrepancies can be directly addressed through consideration of the coupling of lattice degrees of freedom through the standard Franck-Condon formalism. We conclude that similar considerations are necessary in the analysis of $K$-edge RIXS results of more complex transition metal oxides.

\section{Electron-only Model}
With only one hole, the ground state of the single CuO$_4$ plaquette transforms as $x^2-y^2$, with a completely filled 1$s$ electronic core shell, a completely empty Cu 4$p$ shell, and one hole shared among the Cu 3$d$ and O 2$p$ orbitals\cite{tranq91}. A photon with sufficient energy ($\sim$ 8995 eV) can be absorbed, inducing a dipole-allowed optical transition, which results in a 4$p$ electron and a 1$s$ core hole. In the simplest treatment of Cu $K$-edge RIXS, the influence of the core hole is to locally repel valence holes, thereby disturbing the Cu 3$d$ and O 2$p$ valence subsystem, while the extended 4$p$ electron is considered to have a relatively limited influence on the dynamics \cite{tranq91,hanc09, vernay08, vanv93, kotani01}. The RIXS process is completed when the 4$p$ electron radiatively recombines with the 1$s$ core hole, emitting a photon which leaves behind an amount of energy reflective of a core-hole-induced excitation created in the valence system. Because the core hole potential is localized, the resulting spectroscopic process results in electronic transitions among valence states of the same symmetry, so we may limit our attention to states with $x^2-y^2$ character, as is commonly done in considerations of core-level photoemission spectroscopy \cite{vanv93}.

\begin{figure}[htbp]
\begin{center}
\includegraphics[width=5in]{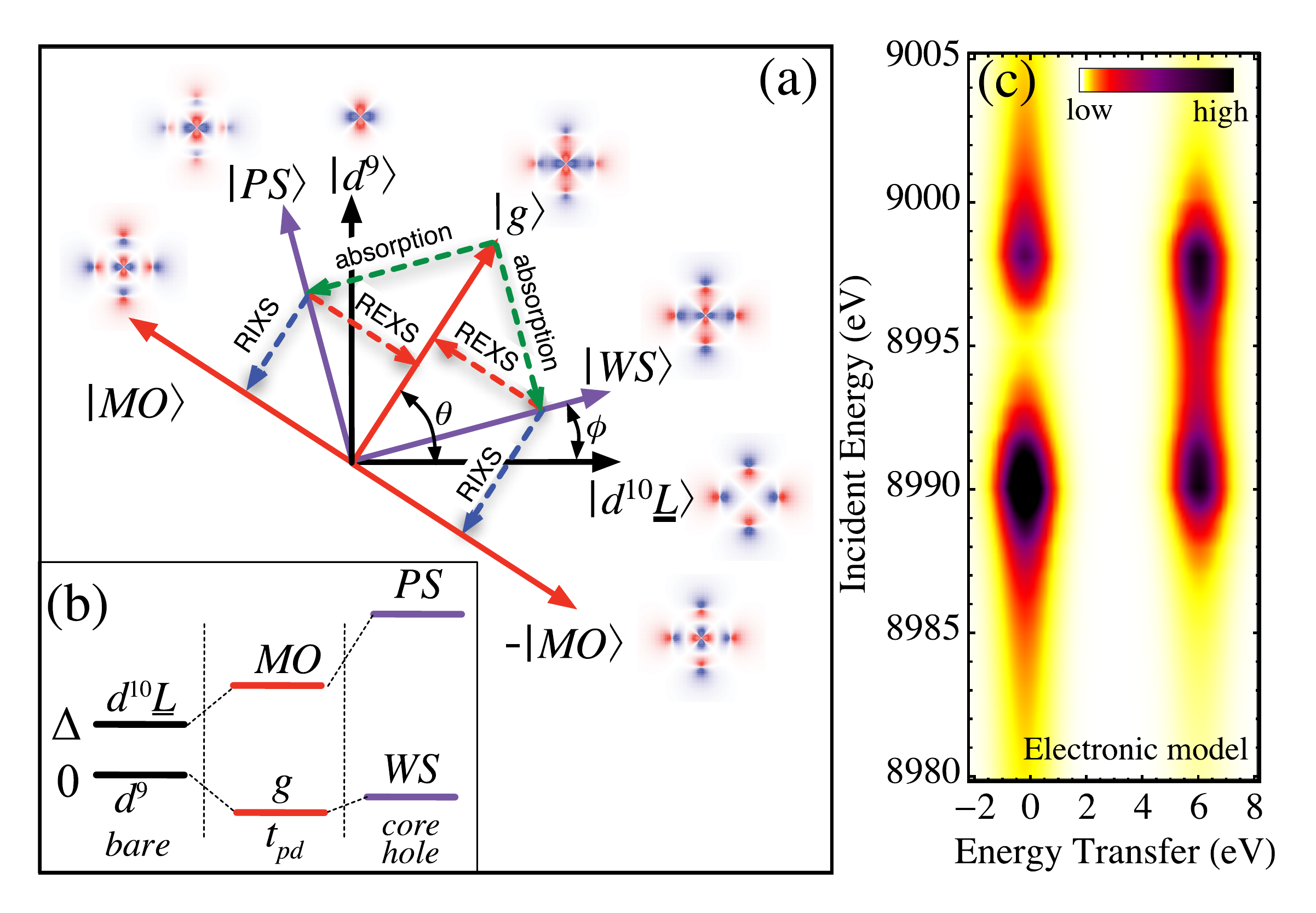}
\caption{(a) Hilbert subspace of a single CuO$_4$ plaquette. Basis states ($d^9$ and $d^{10}\underline{L}$) are shown as black arrows. The effect of introducing hybridization $t_{pd}$ and rediagonalization is effectively a rotation in this picture, giving the red arrows, which show the ground ($g$) and molecular orbital ($MO$) states. Further introduction of the core hole potential gives the purple arrows, depicting the well-screened ($WS$) and poorly screened ($PS$) states. The matrix elements for various spectroscopic processes are represented by dotted lines. Hole wave functions are shown around the perimeter, and the angles $\theta$ and $\phi$ used in the text are defined. (b) Energy level diagram for this system, showing the effect of including $t_{pd}$ and the core hole potential. Colors used correspond to (a). (c) RIXS profile of the single plaquette, after \cite{hanc09}. The state energies are: $E_g$ = 0, $E_{MO}$ = 6 eV, $E_{WS}$ = 8990 eV, $E_{PS}$ = 8998 eV.}
\label{default}
\end{center}
\end{figure}

Following van Veenendaal, Eskes, and Sawatzky \cite{vanv93}, we therefore need only consider the $x^2-y^2$ state $|d^{9}\rangle$ with hole density on the Cu and on the O 2$p$ ligand state of the same symmetry, which we call $|d^{10}\underline{L}\rangle$ (sometimes called $|p_{x^2-y^2}\rangle$ \cite{vanv93}). The wavefunctions for these states are shown in Fig. 1a. These basis states are energetically separated by the bare charge-transfer energy $\Delta$, but are further split under the hybridization $t_{pd}$, and the new ground state is formed as a bonding combination of the two basis states: $|g\rangle=\cos{\theta}|d^{10}\underline{L}\rangle+\sin{\theta}|d^9\rangle$. The antibonding state $|MO\rangle=-\sin{\theta}|d^{10}\underline{L}\rangle+\cos{\theta}|d^9\rangle$ is generally referred to as the molecular orbital (MO) excitation. Transitions from the ground to the MO state are RIXS-active at the Cu $K$ edge because they do not involve changes in symmetry of the electronic wavefunction \cite{hanc09, vernay08, kim04mo}. This is very different from the case of soft X-ray RIXS at the Cu $L_3$ edge \cite{ghir04, ulrich09, braich09, hennies05}, where symmetry-changing $d$$\rightarrow$$d$ transitions are favored over these symmetry-preserving ``charge-transfer" excitations.

During the Cu $K$-edge RIXS process, a deeply bound 1$s$ electron makes a dipole-allowed transition to a spatially extended, high-energy 4$p$ state, leaving behind a 1$s$ core hole. This strong perturbation affects the structure of the valence electronic states, giving rise to two new states, shown as purple arrows in Fig. 1a. The lower-energy ``well-screened" state $|WS\rangle$ = $\cos{\phi}|d^{10}\underline{L}\rangle+\sin{\phi}|d^9\rangle$ has a large electron density screening the core hole potential, while the ``poorly-screened" state $|PS\rangle=-\sin{\phi}|d^{10}\underline{L}\rangle+\cos{\phi}|d^9\rangle$ has a higher hole density. In Cu $L_3$ X-ray photoemission spectroscopy (XPS) \cite{vanv93}, the projection of the ground state onto each of these states determines the spectral profile (e.g. $I_{\rm{XPS}}\propto|\langle WS|g\rangle|^2$). However, in a resonant elastic or inelastic scattering experiment (REXS or RIXS), the signal arises from double projections, first from the ground state ($g$) onto highly excited intermediate states (WS or PS), and then back to the lower-energy excited state (MO). In our case, this double projection has a particular property: $\langle MO|WS\rangle\langle WS|g\rangle=-\langle MO|PS\rangle\langle PS|g\rangle$, independent of $t_{pd}$ or core hole interaction strength, as can be seen graphically in Fig 1a. This fact implies that there is equal strength of RIXS scattering through each of the two intermediate states into the MO final state, leading to a double-peak pattern (Fig. 1c). A streak of intensity between these two resonances occurs due to constructive interference for intermediate photon energies, and is a consequence of the resonant Raman process for two non-degenerate intermediate states with equal and opposite matrix elements \cite{hanc09}.

\section{MO excitation in experiments}
To date, the MO excitation has been observed in cuprates of various electronic dimensionalities: quasi-zero dimensional \cite{hanc09}, edge-shared one-dimensional \cite{vernay08, kim04lico}, corner-shared one-dimensional \cite{hasan02}, two-dimensional planar \cite{abba99, hasan00, kim02, lu05, lu06}, and three dimensional \cite{yuri09}. The electronic structure is far more complicated in the higher-dimensional systems than in the molecular limit, but a localized MO excitation is nonetheless observed \cite{kim04mo}. As discussed above, it appears that this excitation resonates strongly at both the WS and PS states of the low-dimensional systems \cite{hanc09,vernay08}. In general, an approximate correspondence can be made with the WS and PS states discussed above, and it is through a higher-energy PS-like state that the MO excitation resonates most strongly, due to the localized nature of the dominant $d^9$ character. A WS-like intermediate state can also be identified, but because this state involves the displacement of hole density away from the Cu site and onto the neighboring O, it couples most directly to extended excitations of both the charge-transfer \cite{hasan00, abba99, kim02, lu05, lu06} and magnetic \cite{hill08, braich09} types.

Figure 2a shows Cu $K$-edge RIXS data for electronically quasi-zero-dimensional CuB$_2$O$_4$. The data were collected at SPring-8 BL11XU with instrumental energy resolution 380 meV FWHM. The strong Raman-dispersing peak at $\sim$ 7 eV transfer is the MO excitation \cite{hanc09}. The low-energy $d$$\rightarrow$$d$ and higher-energy 4$p$ excitations are identified and discussed in detail elsewhere \cite{hanc09}. Fig 2b shows an anomalous 20\% sharpening of the excitation at an incident energy 9002 eV. Figure 2c shows that there is a clear step-like incident-energy dependence to the MO feature's energy, which mimics the behavior found in the edge-shared cuprate LiCu$_2$O$_2$ \cite{kim04lico}, also shown in Fig. 2c. These observations can not be understood in the electron-only treatment discussed above, and were not reproduced by the purely electronic cluster calculations of Hancock \textit{et al.} \cite{hanc09} or Vernay \textit{et al.} \cite{vernay08}. A third puzzling aspect of the data is the broad Gaussian lineshape, which is unexpected for a long-lived electronic final state that is weakly coupled to its environment. We interpret these salient, but enigmatic properties to imply that purely electronic considerations are insufficient to fully understand the data.

\begin{figure}[htbp]
\begin{center}
\includegraphics[width=5in]{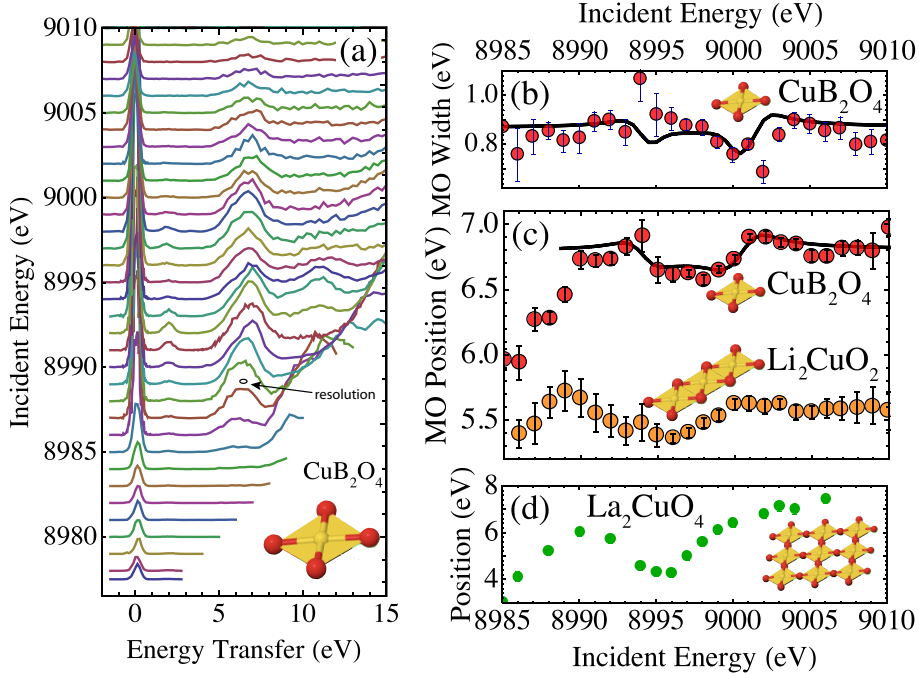}
\caption{(a) Line scans for CuB$_2$O$_4$ \cite{hanc09} showing the incident-energy dependence of the MO excitation. (b) Incident-energy dependence of the MO Gaussian width. The line shows the same for the model discussed in the text, multiplied by a factor of 2.5. (c) Incident-energy dependence of the MO peak position of CuB$_2$O$_4$ (incident polarization $\epsilon_i ||$ $\langle 001 \rangle$) \cite{hanc09} (solid circles) and LiCu$_2$O$_2$ \cite{kim04lico}. The line shows the position from the model calculation (Fig. 3b). (d) The incident-energy dependence of the Gaussian position for La$_2$CuO$_4$ (from \cite{abba99}).}
\label{default}
\end{center}
\end{figure}

\section{Effect of lattice degrees of freedom}

We now make the simplest possible extension of the above model to include the lattice degrees of freedom of the system, and show that these three puzzling aspects of the data can be succinctly addressed. The procedure follows the Franck-Condon (FC) treatment for a molecular system \cite{salek99, hennies05, cast07}, wherein we consider as before transitions among the electronic degrees of freedom ($e$), but with additional coupling to the vibrational continuum ($v$) within the sudden approximation at each stage of the two-step RIXS process. This process is shown schematically in Fig. 3a. 

The RIXS cross section for this model is then given by the Kramers-Heisenberg formula:
\begin{equation}
\frac{d\sigma}{d \Omega}\propto\sum_{m=0}^\infty\big|\mathcal{M}\big|^2 \delta(\hbar(\nu_{in}-\nu_{sc})-(E_{MO}+m_{MO}\hbar\omega_0-E_g)).
\label{ }
\end{equation}
with the amplitude
\begin{equation}
\mathcal{M}=\sum_{n=0}^\infty\sum_{i=WS,PS} \frac{[\langle MO | i\rangle \langle i | g\rangle]_e\hspace{.1in}[\langle m_{MO} | n_{i}\rangle \langle n_{i} | 0_g\rangle]_v}{\hbar\nu_{in}-(E_i^e+\hbar\omega_0 n_{i}-E_g)-i\Gamma}.
\label{ }
\end{equation}
Here $\omega_0$ is the natural frequency of the harmonic potential, $\Gamma$ is the core-hole decay rate, $E_{MO}$, $E_i$, and $E_g$ are the (electronic) energies of the MO, intermediate (WS or PS), and ground states, respectively, and $\hbar\nu_{in}$ ($\hbar\nu_{sc}$) is the energy of the incident (scattered) photon.

The result for the purely electronic model in Fig. 1c is recovered if we assume negligible lattice coupling and molecular potential deformations. As before, we would have a strong MO excitation, resonating equally at the WS and PS states, whose position is given by the bare electronic transition energy $E_{MO}-E_g$, and whose width is narrow, determined by the intrinsic lifetime of the final electronic state. The new elements are the Franck-Condon factors, which represent the overlap of nuclear wavefunctions from the ground and excited state potential energy surfaces. These overlaps can be calculated analytically \cite{cast07} if we assume a purely translational shift of a one-dimensional harmonic potential. The FC factors are then calculated as simple overlaps of harmonic wave functions (Fig. 3a), e.g.:
\begin{equation}
[\langle m_a | n_b\rangle]_v=e^{-\gamma_{a\rightarrow b}^2/2}  \frac{(-1)^{n-N}\sqrt{m!n!}\gamma_{a\rightarrow b}^{n+m-2N}}{(n+m-N)!}L_N^{n+m-2N}(\gamma_{ba}^2)
\label{ }
\end{equation}
where $N=\min{(n,m)}$, $L_N^m$ is the associated Laguerre polynomial, and $\gamma_{a\rightarrow b}=\sqrt{\frac{\mu \omega_0}{2\hbar}}(x_{b}-x_a)$ is the electronic transition-induced displacement of the harmonic potential in units of the ground state positional uncertainty.

The parameters of the model which express the influence of phonon coupling are $\gamma_{g \rightarrow i}$, and $\gamma_{i\rightarrow MO}$, where $i\in\{$WS, PS$\}$, and the phonon frequency $\omega_0$. The latter sets the energy scale for coarse changes in both directions of the energy transfer/incident energy plane. In this plane, the core-hole broadening ($\Gamma$$\sim$1 eV) provides a large electronic energy uncertainty in the incident energy direction. Therefore, the summation over the intermediate state phonon index $n$ has highly overlapping, interfering contributions, with relative phases which are energy dependent close to the resonance. The result of this summation also depends on the final state phonon index $m$, and therefore weakly affects the lineshape in the energy-transfer direction. The model therefore supports incident-energy dependent lineshape properties, with even one electronic transition, in contrast to purely electronic models \cite{hanc09,vernay08}.

\begin{figure}[htbp]
\begin{center}
\includegraphics[width=6in]{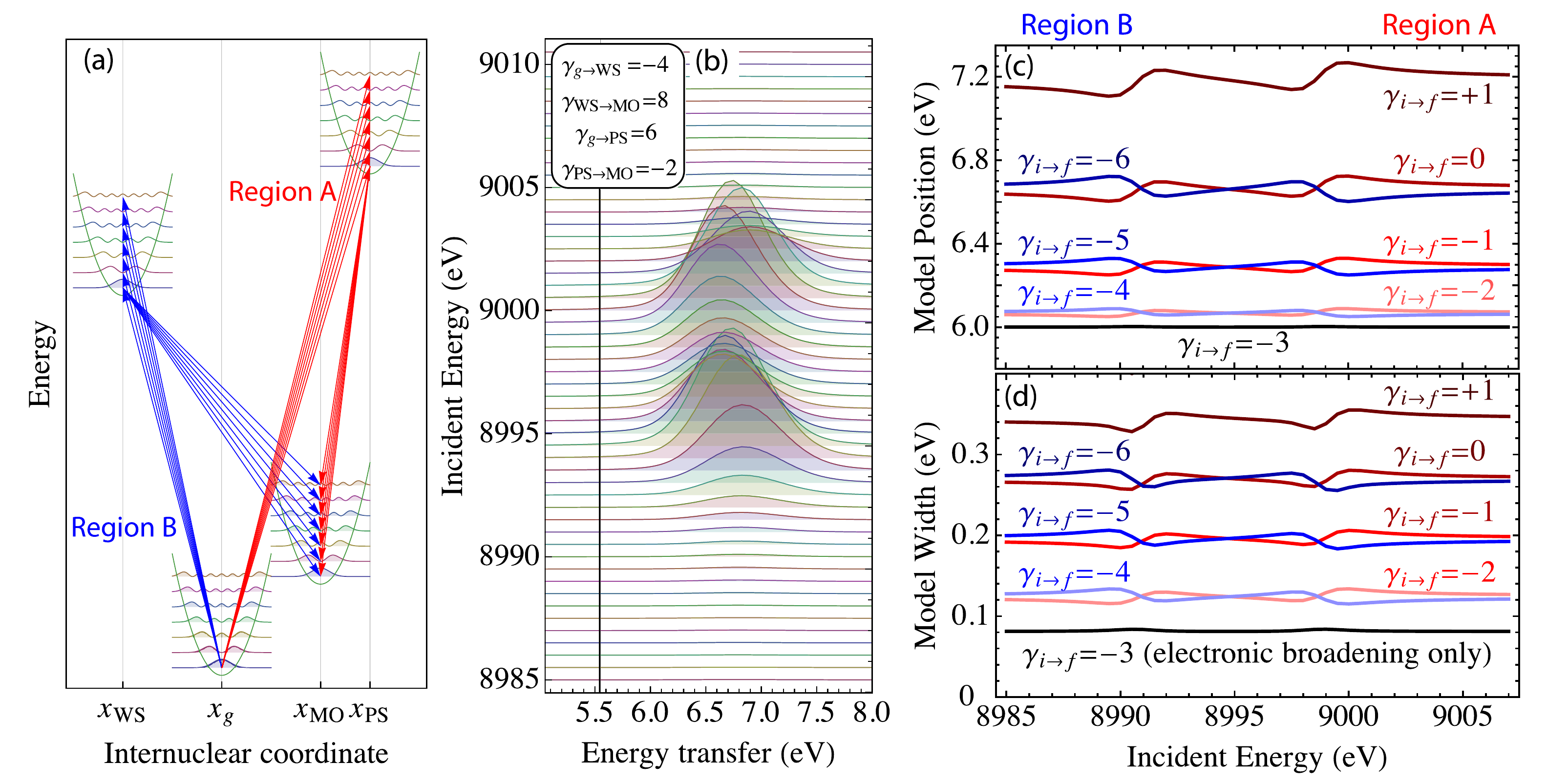}
\caption{(a) Schematic of the Franck-Condon process. The electronic transitions are accompanied by a manifold of vibrational transitions, so that overlaps of the nuclear wavefunctions between ground, intermediate and final states need to be considered. The overlaps strongly depend on the relative displacement of these three states (coordinates $x_g$, $x_{WS}$, $x_{PS}$, and $x_{MO}$), which enter the FC amplitudes as $\gamma_{a\rightarrow b}=\sqrt{\frac{\mu \omega_0}{2\hbar}}(x_{b}-x_a)$. The one-dimensional spatial ordering of these four states is implied by the fit to the model shown in Fig. 2c. (b) Plot of Eq. (1) for the FC factors given in Eq. (2) with $\gamma_{g\rightarrow WS}$ = -4, $\gamma_{WS\rightarrow MO}$ = 8, $\gamma_{g\rightarrow PS}$ = 6, and $\gamma_{PS\rightarrow MO}$ = -2, with $\hbar\omega_0$ = 75 meV. The vertical line at 5.55 eV shows the bare electronic excitation energy. Incident energy-dependent (c) position and (d) width for fixed $\gamma_{g\rightarrow i}=$ +3 and various values of $\gamma_{g\rightarrow f}$. }
\label{default}
\end{center}
\end{figure}

In more detail, the lineshape is dependent on the relationship between $\gamma_{g\rightarrow i}$ and $\gamma_{i\rightarrow MO}$. If $\gamma_{g\rightarrow i}$+$\gamma_{i\rightarrow f}=\gamma_{g\rightarrow f}$=0, there is no net displacement of the potential energy surface and no effect in the energy transfer direction: we again recover Fig. 1c, with an expected narrow Lorentzian. For a fixed $\gamma_{g\rightarrow i}$, moving $\gamma_{i\rightarrow f}$ in either direction away from the symmetry point $\gamma_{g\rightarrow f}$=0, the lineshape blueshifts and quickly becomes Gaussian for even the modest value of $|\gamma_{g\rightarrow f}|\sim$ 3. Figs. 3c and 3d respectively show the position and width for a range of values of $\gamma_{i\rightarrow f}$ with $\gamma_{g\rightarrow i}$ = +3. In addition to the blueshift of order $\hbar$$\omega_0$$\gamma_{g\rightarrow f}^2$ and a broadening of order $\hbar\omega_0$$|\gamma_{g\rightarrow f}|$, step-like modulations appear in the energy position and width when the incident photon energy passes through resonance. The direction of the step in incident energy position depends on whether 0 $<\gamma_{g\rightarrow f}$$\gamma_{g\rightarrow i}$ (region A) or 0 $>\gamma_{g\rightarrow f}$$\gamma_{g\rightarrow i}$ (region B). These two limits are sketched in Fig 3a as a set of red or blue arrows, respectively.

The comparison in Fig. 2b of these trends in the incident-energy-dependent position in the data for CuB$_2$O$_4$ and the calculation suggests that the PS state is in region A. This assignment is consistent with the fact that there is an abrupt stiffening and narrowing of the feature as $\hbar\nu_{in}$ is passes through the PS resonance around 8998 eV. The situation for the WS state is less clear, due to the lack of narrowing at lower incident energies, as well as likely interference from the approaching $K\beta_{2,5}$ fluorescence line, which is a relatively strong 3$d$$\rightarrow$1$s$ radiative transition at fixed scattered photon energy 8976 eV (Fig. 2a). Despite ambiguities regarding the description of the WS state, we tentatively put the WS state in region B based on the step in the MO position near 8994 eV. Fig. 3b shows the model results with $E_{PS}-E_{WS}$=5.55 eV and ($\gamma_{g\rightarrow WS}$, $\gamma_{WS\rightarrow MO}$, $\gamma_{g\rightarrow PS}$, $\gamma_{PS\rightarrow MO}$) = (-4, 8, 6, -2), corresponding to $\gamma_{g\rightarrow f}$=4. The vibrational energy was chosen to be $\hbar\omega_0$ = 75 meV, which is a typical value for bond-stretching phonons in cuprates \cite{mang04}. The values of the four parameters $\gamma$ imply a spatial ordering to the equilibrium positions shown in the schematic of Fig. 3a.

Having established that the phonon-coupled model in Eqns. ((1)-(3)) is capable of describing the data at a qualitative level, we point out limitations that may be important to a quantitative application. First, the experimental value of the MO width ($\sim$ 760 meV) is about 2.5 times larger than the model value. The quantitative comparison has several potential shortcomings. One is the large uncertainty in the intrinsic width of the electronic excitation, which could strongly effect a quantitative analysis based on the widths of the MO feature. If we assume a larger intrinsic width, we can partially remedy the discrepancy, but still underestimate the narrowing near $\hbar\nu_{in}$ = 9000 eV.
Another point is that the curvature of the internuclear potential energy is not likely to be the same at each stage of the process, as we have assumed. Such effects would manifest themselves as different values of $\omega_0$ for the g, MO, WS, and PS states. In an extreme case, the potential energy surfaces may not even support bound vibrational states, but rather have an unstable, dissociative nature. Additional complications arise from the fact that CuB$_2$O$_4$ contains two crystallographically distinct plaquette types, which may inhomogeneously broaden the lineshape, leading to an overestimate of the importance of phonons based on the width alone. Furthermore, there exist scattering channels which are not included in the above considerations, such as those which give rise to the $d$$\rightarrow$$d$ and 4$p$ transitions described in \cite{hanc09}. It is therefore possible that the peak is composed of several electronic transitions, further obfuscating an interpretation of its width. While a detailed quantitative application of the FC model with quantum chemistry methods is beyond the scope of the present work, we have shown that this model has the desired properties to describe the Gaussian lineshape with concurrent resonance-induced narrowing and step in energy transfer for the PS state.

\section{Implications, conclusions, and outlook}

We now discuss some general implications of our observations. Kim \textit{et al.} \cite{kim04mo} performed a systematic study of the molecular orbital excitation energy for a wide range of cuprates, and obtained the surprising result that this energy scales with the copper-oxygen bond length $d_{\mathrm{Cu-O}}$ as $d_{\mathrm{Cu-O}}^{-\alpha}$ with $\alpha\sim$ 8$\pm$2. This behavior is very different from that expected based on Harrison's rules \cite{harrbook} and band structure calculations \cite{cooper90}, for which the exponent is in the range $\alpha$$\sim$3.5-4. We identify the feature observed in RIXS as the maximum of a Franck-Condon band (at energy $E_{MO}+\hbar\omega_0|\gamma_{g\rightarrow f}|-E_g$), and not as the bare electronic transition (zero-phonon line at $E_{MO}-E_g$), which would reside at much lower energy (e.g., the vertical line in Fig. 3b). Consequently, in addition to a purely electronic contribution with small exponent, we expect a significant additive contribution that arises from the bond-length ($d_{\mathrm{Cu-O}}$) dependence of the molecular potential shifts ($\gamma_{i\rightarrow f}$ and $\gamma_{g\rightarrow i}$). In our interpretation, this dependence is determined by the way the relevant phonon frequencies and potential shifts depend on the bond length. Our result suggests that the parameter combination $\hbar\omega_0$$|\gamma_{g\rightarrow f}|$ behaves approximately as $d_{\mathrm{Cu-O}}^{-8\mp2}$.

Reference \cite{hanc09} assigns the weak $\sim$2 eV of CuB$_2$O$_4$ (reproduced in Fig. 2a) to a $d$$\rightarrow$$d$ excitation and showed that this transition is a robust, one-plaquette effect, and not the result of local symmetry-lowering distortions. This transition is anomalous, because it necessarily involves a change in symmetry, an effect unexpected at the Cu $K$ edge. In the context of phonon coupling, we can address its appearance by considering the effect of the 4$p$ electron on the molecular potential surface. If the photoelectron reaches, say, a 4$p_x$ state, anisotropic screening could force the magnitude of molecular displacement of the O placed along the $\pm x$ axis to be different from the displacement of the O placed along the $\pm y$ axis. This distortion of the plaquette in the $B_{1g}$ channel in the intermediate state could open a pathway to a $d_{3z^2-r^2}$ final state, for example. In Ref. \cite{hanc09}, it was found that the $d$$\rightarrow$$d$ excitation resonates more strongly at the WS and PS states with 4$p_{x,y}$ polarization than with 4$p_z$ polarization, thus supporting this scenario for $d$$\rightarrow$$d$ genesis. Future high-resolution data for CuB$_2$O$_4$ or other low-dimensional cuprate systems may help give better insight into the nature of the $d$$\rightarrow$$d$ transition and the plausibility of this scenario.

Finally, we discuss the possible extension of the lattice-coupled RIXS cross section to planar cuprates. Abbamonte \textit{et al.} \cite{abba99} reported a 6 eV excitation at the $K$ edge of Cu in La$_2$CuO$_4$, which was described by a Gaussian line shape dressed by resonant denominators representing the incident and scattered photon poles. Later work in a different scattering geometry \cite{kim02,lu05,lu06} resolved a fine structure of excitations, so the original work in Ref. \cite{abba99} should be regarded a measure of the overall inelastic spectral weight, containing both the MO excitation as well as charge-transfer excitations. As seen from Fig. 2, the incident-energy-dependent position of this spectral weight exhibits qualitative similarities with the MO excitation results for CuB$_2$O$_4$ and LiCu$_2$O$_2$ as well as our model calculation. The magnitude of the shift is much larger for La$_2$CuO$_4$, perhaps reflecting the enhanced range of influence of the core hole potential in generating phonons, or additional effects due to the complex electronic structure of the CuO$_2$ plane. More recent experimentation \cite{abba99, hasan00, kim02, lu06, ellis08} revealed that a sharp, relatively weak 2.2 eV charge-transfer feature resonates at the WS state of La$_2$CuO$_4$ and appears to have a temperature-dependent lineshape \cite{ellis08}. The latter observation may be addressed already within a simple extension of our model: we have considered transitions only from the electronic-vibrational ground state, but transitions from the ensemble of thermally activated vibrational states would inevitably change the line profile, suggesting a route to understanding the observed temperature dependence at the relatively high energy of $\sim$ 2.2 eV.

In conclusion, we have shown that the electron-only models are incapable of describing even salient experimental results at the transition metal $K$-edge. On the other hand, we demonstrate for the molecular-orbital excitation of the simple model system CuB$_2$O$_4$ that considerations of the coupling of the lattice to electronic degrees of freedom leads to a natural interpretation of the observed line shapes and (incident) photon-energy dependence. Nonadiabatic effects were first observed in C $K$-edge RIXS of ethylene \cite{hennies05}, and our results show that these effects carry to the 3$d$ transition metal edges as well. Phonon coupling is likely present in existing RIXS data for cuprates \cite{lu05, lu06, hasan00} and other transition metal oxides \cite{ulrich09, grenier05}, as we have shown for LiCu$_2$O$_2$ \cite{kim04mo} and La$_2$CuO$_4$ \cite{abba99}. We have also identified an important property of the model that the lineshape away from resonance is most heavily influenced by the electron-phonon coupling between the ground and final state-it is only near resonance that the intermediate state effects play a role. 
Finally, we emphasize that the process discussed here is only one possible channel for RIXS scattering, where the photoelectron recombines, and the charge excitations occur only via the core hole interaction. In the case of $d$$\rightarrow$$d$ excitations \cite{ghir04}, orbiton scattering \cite{ulrich09}, and magnetic \cite{hill08, braich09} excitations, the applicability of the present considerations is not as clear. We expect future experiments with higher energy resolution, coupled with more refined theoretical considerations, to directly address the role of the electron-phonon interaction in determining RIXS lineshapes.

We would like to acknowledge valuable conversations with Girsh Blumberg, Dirk van der Marel, and Fran\c{c}ois Vernay. JNH would like to acknowledge the continuing support of Eleiza Braun.
This work was supported by the DOE under Contract No. DE-AC02-76SF00515.

\end{document}